\documentclass[11pt]{article}
\makeatletter\@addtoreset {equation}{section}\makeatother

\usepackage[dvips]{epsfig}
\usepackage{graphicx}

\usepackage{amsmath}
\usepackage{amsfonts,amssymb}

\newcommand{\RM}{\mathbb{R}}
\newcommand{\CM}{\mathbb{C}}

\newcommand{\sech}{\mathrm{sech}}

\begin{document}

\bibliographystyle{plain}
\nocite{*}

\title{\bf Solitary Waves Under the Competition of Linear and Nonlinear 
Periodic Potentials}
\author{Z. Rapti$^{1}$, P.G. Kevrekidis$^{2}$, V.V. Konotop$^3$, 
C.K.R.T. Jones$^4$ \\
{\small $^{1}$ Department of Mathematics
University of Illinois at Urbana-Champaign,  Urbana, Illinois 61801-2975} \\
{\small $^{2}$ Department of Mathematics and
Statistics, University of Massachusetts, Amherst, MA 01003 } \\
{\small $^{3}$ Centro de F\'{\i}sica Te\'{o}rica e Computacional, Universidade de Lisboa,
  Av. Prof. Gama Pinto 2, Lisboa 1649-003,} 
  \\ {\small
    Portugal and 
  Departamento de F\'{\i}sica, Faculdade de Ci\^encias,
Universidade de Lisboa, Campo Grande,
} \\ {\small 
  Ed. C8, Piso 6, Lisboa
1749-016, Portugal.} \\
{\small $^{4}$ Department of Mathematics
University of North Carolina-Chapel Hill
Chapel Hill, NC 27599-3250} \\
}
\date{\today}
\maketitle

\begin{abstract}
In this paper, we study the competition of linear and nonlinear
lattices and its effects on the stability and dynamics of bright solitary
waves. We consider both lattices in a perturbative framework, whereby
the technique of Hamiltonian perturbation theory 
can be used to obtain information about the existence of 
solutions, and the same approach, as well as eigenvalue count
considerations, can be used to obtained detailed conditions about
their linear stability. We find that the analytical results
are in very good agreement with our numerical findings and can also
be used to predict features of the dynamical evolution of such solutions.
\end{abstract}

\section{Introduction}

The field of Bose-Einstein condensates (BECs) in atomic 
physics \cite{dalfovo} has provided a renewal of interest in solitary
wave and nonlinear excitations. This is especially so because the
leading order, mean-field description of 
interatomic interactions corresponds
to a cubic nonlinearity. This, in turn, is responsible for
the emergence and experimental observation of a variety
of matter-wave solitons in BECs, including bright  \cite{expb1,expb2},
dark \cite{dark} and gap \cite{gap} solitons. Furthermore, 
atom optical devices have been proposed, such as, for instance,
the atom chip \cite{Folman} that would allow the possibility
to controllably manipulate such nonlinear structures. 

It is well-known that bright (respectively, dark) matter-wave 
solitons arise in BECs with attractive (respectively, repulsive)  
interatomic interactions, i.e., for atomic species with 
negative (respectively, positive) scattering length $a$.

One of the particularly appealing features of the BEC setting
is the existence of a wide variety of experimental ``knobs''
that can be used to manipulate or control the relevant structures.
In particular, interfering laser beams can be used to produce
a standing wave pattern, known as the optical lattice, providing
a periodic linear potential for the condensate. This type of structure
offers a large variety of interesting phenomenology including
Bloch oscillations, Landau-Zener tunneling, dynamical instabilities,
gap excitations among many others and a considerable amount of
review works have already been dedicated to this topic 
\cite{pgk,konotop,morsch}. 

On the other hand, 
magnetically-induced Feshbach resonances can be used
to modify at will both the magnitude and the sign of the
scattering length by tuning the external magnetic field; 
see  e.g.  \cite{feshbachNa}
and also \cite{expb1,expb2} where the Feshbach resonance in 
$^{7}$Li BECs was used 
for the formation of bright matter-wave solitons.
The ability to modulate the scattering length has
led to a large variety of studies where this mechanism
has been used. For instance, time-dependent modulations
of the scattering length were proposed as a means of
preventing collapse in higher-dimensional BECs
\cite{FRM1}, or as a way of producing robust matter-wave
breathers \cite{FRM2}, among others. 

A more recent suggestion has been to add to a constant bias magnetic field
a gradient in the vicinity 
of a Feshbach resonance, allowing for a spatial variation of the
scattering length, thereby providing what has come to be 
termed a ``collisionally inhomogeneous environment''.
Notice that, given the availability of magnetic
and optical (laser-) fields, the external trapping potential
and the spatial variation of the scattering length can
be adjusted independently (see \cite{our2} for more
details on the relevant configuration). 
In this latter setting, a variety of propositions
of interesting dynamical phenomena have been made 
concerning scenarios for the emission of solitons \cite{vpg12}, delocalizing transition of matter waves~\cite{LocDeloc},
or the dynamics of the waves in random \cite{vpg14},
linear \cite{vpg15}, periodic \cite{vpg16,BludKon} or localized
\cite{vpg17} spatial modulations. A number of more mathematically
minded results on the existence and stability of waves
have also appeared in \cite{vpg18} and a technique for
analytically constructing exact solutions in \cite{vpg19}.

The framework of collisionally inhomogeneous environments
in combination with external optical lattices provides
an ideal environment for competition. This was illustrated in examples of 
modulational instability of Bloch states~\cite{BludKon} and of the 
delocalizing transition in one-dimension~\cite{LocDeloc}. In the 
present work we develop these ideas exploring 
more general lattice profiles.  In particular, 
external (linear) potentials and collisional
(nonlinear) potentials in (as well as out of) phase
will be considered. It will be demonstrated that
when in phase, these potentials provide a competition
leading
to a number of interesting effects including
stabilization/destabilization thresholds and even
the mutual annihilation of the two potentials to
provide an effectively quasi-translationally-invariant
environment. The effective potential landscape where
the solitary waves (of the bright type) live will
be obtained following the Lyapunov-Schmidt considerations
of \cite{todd1}. Then, the relevant (translational) 
eigenvalue of the linearization will be computed based
on the curvature of this effective potential landscape
and the stability/instability of the waves will be
assessed (and the relevant transition points will be
obtained). These results will be confirmed by a second
independent method based on a direct count of unstable
eigenvalues. 
Finally, we will examine the variation of
a number of relevant key parameters (such as the
amplitude, the wavenumbers or the relative phase)
of the potentials, in order to evaluate the validity
of the approach. Furthermore, when the structures
are unstable, we will examine what this approach
can suggest regarding the actual instability evolution 
dynamics. The understanding that we will develop
will enable us to manipulate the ensuing solitary
waves in such a complex territory and to understand
their dynamical behavior in the presence of linear
and nonlinear lattices.

Our presentation will be structured as follows.
In section II, we will present our analytical
results; in section III, we will corroborate
these results by means of numerical
computations. Finally, in section IV, we will
summarize our findings and present some interesting
directions for future study.

\section{Analytical Results: Solitary Wave Statics and Dynamics}

The prototypical 
framework in which we will consider the above discussed competition
of linear and nonlinear lattices is that of the 
perturbed nonlinear Schr\"odinger equation of the form
\begin{eqnarray}
i u_t =-\frac{1}{2} u_{xx} - \left(1+\epsilon n_1(x) \right) |u|^2 u 
+\epsilon n_2(x)u.
\label{pnls}
\end{eqnarray}
In Eq. (\ref{pnls}), $(x,t)\in \RM \times \RM^+$ and $u \in \CM$. 
While we will keep the presentation of the mathematical results
as general as possible, the particular case of interest in the selection
of the nonlinear and linear lattice will, respectively, be:
\begin{eqnarray}
\begin{array}{l}
n_1(x)=A \cos(k_1 x)
\\ 
n_2(x)=B \cos(k_2 x +\Delta \phi).
\end{array}\label{pots}
\end{eqnarray}
where $A$, $B$, $k_1$, $k_2$ and $\Delta \phi$ are real constants.
Notice that the lattices have the same functional form, which will 
allow us to reveal more lucidly the relevant competition between
the corresponding terms.

When $\epsilon =0$, Eq. (\ref{pnls}) has the well-known 
stable localized soliton solution given by
\begin{eqnarray}
u(x,t)=\sqrt{\mu}\sech [\sqrt{\mu} (x-\xi)] e^{i [v(x-\xi)+\delta]}
\label{soliton}
\end{eqnarray}
where $\mu >0$, $\xi=vt$ is the position of the soliton center, $v$ is the 
velocity of the soliton, and $\delta=(v^2+\mu)/2$.


We presently focus on the stationary modes with $v=0$.
Given the monoparametric nature of the family of the respective soltuions, we
can fix $\mu$ in what follows (in fact, we will fix $\mu=2$ in our
numerical computations below).
Because of the rotational and translational invariance of the unperturbed 
equation, this solution is unique only up to rotational and translational 
symmetry. 

On the other hand, when $\epsilon >0$, the translational invariance of 
the equation is
broken, which may naturally lead to the potential 
destabilization of the localized states, depending on the 
perturbation parameters. This is the problem that we will examine
in what follows under the influence of both 
linear and nonlinear lattices.

\subsection{Hamiltonian Perturbation Approach}

The existence and nature of localized solutions to perturbed Hamiltonian
systems, of which Eq. (\ref{pnls}) is a particular case, was studied  
in \cite{todd1} (and subsequently in a broader setting in
\cite{kks}). A general perturbative approach was developed in these
works based
on Lyapunov-Schmidt solvability conditions \cite{golub}, and relevant
stability calculations were formulated on the basis of the 
Evans function \cite{agj,kap}. Here, we present some of the general features
of the theory, adapt our problem to the general framework of
\cite{todd1,kks} and subsequently apply these methods to the problem
of interest.
In order to apply these criteria, it is convenient to recast Eq. 
(\ref{pnls})  as
\begin{eqnarray}
\frac{du}{dt}=-i \frac{\delta E}{\delta u^*}, 
\label{pnls0}
\end{eqnarray}
where $E(u)=E_0(u)+\epsilon E_1(u)$. Here,
\begin{eqnarray}
E_0(u)=\int_{-\infty}^{+\infty} \frac{1}{2} \left(|u_x|^2-|u|^4\right) dx,
\end{eqnarray}
and
\begin{eqnarray}
E_1(u)=\int_{-\infty}^{+\infty} \left(n_2(x)|u|^2- \frac{1}{2} n_1(x)|u|^4\right) dx.
\end{eqnarray}

Then, for fixed $\mu$, the intuitive condition for the
persistence of the wave is given by \cite{todd1} 
\begin{eqnarray}
\nabla_{\xi} E_1(u)=0,
\label{extra1}
\end{eqnarray}  
where $\xi$ is the previously-free 
parameter associated with the invariance (in the
case of translation, it is the center of the pulse (\ref{soliton})). 
This condition implies that the wave is going to
persist only if centered at the parameter-selected 
extrema of the energy (which are now going to form, at best, a countably
infinite set of solutions, as opposed to the one-parameter infinity of 
solutions previously allowed by the translational invariance).

Equally importantly, from this expression and from the nature of the
wave, one can infer stability information about the solution of interest.
In particular, the stability of the perturbed wave is determined by
the location of the eigenvalues associated with the translational
invariance; previously, the relevant 
eigenvalue pair was located at the origin $\lambda=0$
of the spectral plane of eigenvalues $\lambda=\lambda_r + i \lambda_i$.
On the other hand, we expect the eigenvalues associated with the
$U(1)$ invariance (i.e., the phase invariance associated with the
$L^2$ conservation) to remain at the origin, given the preservation
of the latter symmetry under the perturbations considered herein.
To compute the relevant eigenvalues we refer to the framework
put forth by the works of \cite{todd1,kks} (adapting the notation
of the latter work) in the following form. Using Proposition 6.1 of
\cite{kks}, we expect that the perturbed system eigenvalues will
be given by the matrix equation:
\begin{eqnarray}
{\rm det}[M_1 + \lambda_1^2 D_G]=0
\label{new1}
\end{eqnarray}
where
\begin{eqnarray}
D_G=
\left( \begin{array}{cc}
(\partial_x u^0,-x u^0) & 0 \\
0  & (u^0,\partial_{\mu} u^0) \end{array} \right)
= \left( \begin{array}{cc}
\mu^{1/2} & 0 \\
0  & -\mu^{1/2} \end{array} \right);
\label{new2}
\end{eqnarray}
also, $M_1$ is given by 
\begin{eqnarray}
M_1=
\left( \begin{array}{cc}
\frac{\partial}{\partial \xi}(\frac{\partial E_1}{\partial (u^0)^{\star}}, \partial_{\xi} u^0) & 0 \\
0  & 0 \end{array} \right)
=\left( \begin{array}{cc}
\int \left( \frac{1}{2} \frac{d^2 n_2}{d x^2} (u^0)^2 - \frac{1}{4}  
\frac{d^2 n_1}{d x^2} (u^0)^4 \right) dx & 0 \\
0 & 0 
\end{array} \right).
\end{eqnarray}
In the formulation of \cite{kks}, the relevant eigenvalues
are obtained to leading order as $\lambda=\sqrt{\epsilon} \lambda_1$, and $u^0$
denotes the solitary wave of Eq. (\ref{soliton}).
Hence, we
conclude from Eq. (\ref{new1}) that as indicated above,
the eigenvalues associated with the rotational invariance
will be preserved at $\lambda=0$, while the translational
eigenvalue will be shifted according to:
\begin{eqnarray}
\lambda^2=- \frac{\epsilon}{\mu^{1/2}} 
\int \left( \frac{1}{2} \frac{d^2 n_2}{d x^2} (u^0)^2 - \frac{1}{4}  
\frac{d^2 n_1}{d x^2} (u^0)^4 \right) dx
\label{extra2}
\end{eqnarray}

Hence, the corresponding
eigenvalue can be straightforwardly evaluated, provided that we
first compute the extrema of the effective energy landscape 
$E_1$, which, as a function of $\xi$, will hereafter be denoted
as $V_{eff}(\xi)$, or more precisely $V_{eff}(\xi)=\epsilon E_1$. 
This $V_{eff}$ will be the effective energy landscape and the
stability or instability of the configuration will be associated
with the convexity or concavity of this effective energy landscape.

We now proceed to evaluate the relevant expressions of the general
theory for the special case of interest herein, namely for the
potentials of Eq. (\ref{pots}).
The effective energy landscape can be evaluated after performing two
straightforward contour integrations, that yield
\begin{eqnarray}
V_{eff}(\xi)=-\epsilon \pi A k_1 \left(k_1^2+4 \mu \right) 
\frac{\cos(k_1 \xi)}{12 \sinh \left(\frac{\pi k_1}{2 \sqrt{\mu}}\right)}
+ \epsilon \pi B k_2 
\frac{\cos(k_2 \xi + \Delta \phi)}{\sinh \left(\frac{\pi k_2}{2 \sqrt{\mu}}
\right)}
\label{cintgeneral}
\end{eqnarray} 
In the simple case of $A=\Delta \phi=0$, the above result reduces to that
in \cite{todd1}, which leads to the well-known conclusion of \cite{oh},
according to which a maximum of a linear
periodic potential leads to an unstable
solitary wave configuration, while the opposite is true for a minimum of
a periodic potential. However, in our case, there is an intriguing 
interplay between the $B$-dependent term stemming from the linear
optical lattice and the $A$-dependent term, emerging from the nonlinear
optical lattice. This competition leads to the potential for 
stability-instability transitions for the wave, based on the properties
of the trapping (such as $(A,B,k_1,k_2,\Delta \phi)$), but also
the properties of the wave itself (since the expression of 
(\ref{cintgeneral}) is explicitly dependent on $\mu$).

\subsection{Eigenvalue Count Approach}

It is worthwhile to note, however, that the above stability results, based
on the formulation of \cite{todd1,kks} can also be 
alternatively derived using the 
approach of \cite{oh,jones}. We present this alternative formulation here,
since we consider it to be a nice complement to the direct eigenvalue
computation of \cite{todd1,kks} using the Hamiltonian
perturbation  technique.
It is well-known that the stability of the solitary wave \cite{kks}
is determined by the number of negative eigenvalues of the operators
\begin{eqnarray}
L_-^{\epsilon} &=& -\frac{1}{2} \frac{d^2}{d x^2} - (1+ \epsilon n_1) 
(u^{\epsilon})^2  + \epsilon n_2 + \frac{\mu}{2}
\label{l-}
\\
L_+^{\epsilon} &=& -\frac{1}{2} \frac{d^2}{d x^2} - 3 (1+ \epsilon n_1) 
(u^{\epsilon})^2  + \epsilon n_2 + \frac{\mu}{2}
\label{l+}
\end{eqnarray}
In particular, if $n(L)$ denotes the count of negative eigenvalues 
and $|n(L_+^{\epsilon})-n(L_-^{\epsilon})| > 1$ \cite{jones}, then the 
solitary wave is unstable. The superscript $\epsilon$ in the operators is
to distinguish the $\epsilon \neq 0$ and the $\epsilon=0$ cases.
In the latter, $L_+^0$ is well known to have a single negative 
eigenvalue ($-3 \mu/2$) with an eigenfunction spanned by $(u^0)^2$
and a single zero eigenvalue with an eigenfunction spanned by
$du^0/dx$, while $L_-^0$ has no negative eigenvalues and a single
zero eigenvalue with an eigenvector spanned by $u^0$. In the
perturbed case, $L_-^{\epsilon}$ retains its zero eigenvalue with
eigenvector spanned by $u^{\epsilon}$. 
Also, from the perturbation theory of Schr\"odinger operators \cite{kato} it
is known that $L_+^{\epsilon}$ has a negative eigenvalue near $-3 \mu/2$,
and a second eigenvalue near $0$, which will be denoted by 
$\lambda_{L_+^{\epsilon}}$. 
Both of these 
eigenvalues are analytic in $\epsilon$, at least in a neighborhood of the
real axis.
Then the stability question
(given also that for this branch of solutions $dN/d\mu>0$ \cite{kks})
is completely settled by the count of negative eigenvalues of 
$L_+^{\epsilon}$. In particular:
\begin{itemize}
\item If $L_+^{\epsilon}$ has two negative eigenvalues, then the
solitary wave will be unstable since 
$|n(L_+^{\epsilon})-n(L_-^{\epsilon})|=2$;
\item If $L_+^{\epsilon}$ has only one negative eigenvalue, then the
coherent structure will be stable.
\end{itemize}
Hence, the stability issue hinges on the shift of the zero eigenvalue
(corresponding to translational invariance, when $\epsilon=0$) in
the presence of the perturbation. Thus, similarly to \cite{oh}, we will
consider the quantity $(L_+^{\epsilon} u, u)$ with $u$ being the eigenvector
corresponding to the eigenvalue $\lambda_{L_+^{\epsilon}}$; here
$(v,w)$ denotes the $L^2$-inner product of $v$ with $w$.
$u$ can be decomposed as $u=u_{\parallel} + u_{\perp}$, where $u_{\parallel}$ 
is proportional to $d u^{\epsilon}/dx$ and $(u_\perp,d u^{\epsilon}/dx)=0$. 
Along the lines of \cite{floer, oh2} one can show that 
$||u^{\epsilon}-u^0||_{H^2}\to 0$, and therefore 
$u_{\parallel} \rightarrow d u^0/dx$
(up to a proportionality factor) as $\epsilon \rightarrow 0$. 
From perturbation theory \cite{funan} it is also known that the eigenvector
$u$ of
$L_+^{\epsilon}$ is analytic in $\epsilon$. Therefore 
$u_{\perp} \rightarrow 0$, 
and actually $u_{\perp} \sim \epsilon$, i.e., it will be of the order of the 
perturbation. Then, we have
\begin{eqnarray}
(L_+^{\epsilon} u , u)= (L_+^{\epsilon} u_{\parallel}, u_{\parallel})
 + 2 (L_+^{\epsilon} u_{\parallel} , u_{\perp}) + (L_+^{\epsilon} u_{\perp} ,
u_{\perp}).
\label{interm}
\end{eqnarray}
However, each of the second and third terms will be of order higher
than the first (at least O$(\epsilon^2)$, while the dominant one will
be of O$(\epsilon)$), hence the solitary wave stability will be determined
by $ (L_+^{\epsilon} u_{\parallel}, u_{\parallel})$. But then,
\begin{eqnarray}
(L_+^{\epsilon} u , u)= r^2 (L_+^{\epsilon} \frac{d u^{\epsilon}}{d x} ,
 \frac{d u^{\epsilon}}{d x} ) + {\rm O}(\epsilon^2);
\label{interm2}
\end{eqnarray}
$r$ is an appropriate proportionality factor (between $u_{\parallel}$
and $d u^{\epsilon}/dx$).
By means of a direct computation (differentiating the equation
satisfied by the stationary state) one has that 
\begin{eqnarray}
L_+^{\epsilon} \frac{d u^{\epsilon}}{d x} = - \epsilon \frac{d n_2}{dx} 
u^{\epsilon} + \epsilon \frac{d n_1}{dx} (u^{\epsilon})^3,
\label{interm3}
\end{eqnarray}
 which, in turn, forming the inner product with $d u^{\epsilon}/dx$,
and integrating by parts leads to the key  result, namely:
\begin{eqnarray}
(L_+^{\epsilon} u , u)= r^2 \epsilon 
\int \left( \frac{1}{2} \frac{d^2 n_2}{d x^2} (u^0)^2 - \frac{1}{4}  
\frac{d^2 n_1}{d x^2} (u^0)^4 \right) dx + {\rm O}(\epsilon^2)
\label{interm4}
\end{eqnarray}
The relevant integral term 
of the right hand side can be seen by direct inspection
to be equivalent (up to a negative-definite proportionality factor) 
to the expression for $\lambda^2$ in Eq. (\ref{extra2}). 
Its positivity (indicating a shift of the zero
eigenvalue to positive values) will imply stability, while its negativity
(indicating a shift of the zero eigenvalue to negative values) will 
lead to instability. This conclusion is fully equivalent to the ones
obtained from Eq. (\ref{extra2}) [although the latter, in some sense,
contains additional information yielding a quantitative measure of
the relevant eigenvalue].

\subsection{Solitary Wave Dynamics}

In order to describe the dynamics of a soliton of Eq. (\ref{pnls}) at 
$\epsilon\ll 1$ one can employ 
the perturbation theory for the NLS soliton~\cite{KarpmanMaslov}, or more precisely, the adiabatic approximation. If 
$\epsilon\neq 0$ then $\xi\neq vt$ 
and has to be found from the equations of the adiabatic approximations. 
The straightforward algebra yields
\begin{eqnarray}
\label{adiabatic}
\frac{d^2\xi}{dt^2}=-\frac 1N \frac{\partial V_{eff}(\xi)}{\partial \xi}
\end{eqnarray}
where $N=\int |u|^2dx=2\sqrt{\mu}$ is the number of particles [it is an 
integral of motion of (\ref{pnls})].

It follows from (\ref{adiabatic}) that there exist different types of 
motion of the soliton. In 
particular if $k_2=k_1=k$ and $\Delta\phi=0$ the soliton dynamics 
reproduces the mathematical 
pendulum. The respective motion of a soliton can be either periodic or 
translational (i.e., unbounded), depending 
on the initial conditions. Another special case arises for
 $A(k^2+4\mu)=12 B$, in which case the right hand side 
of (\ref{adiabatic}) 
becomes zero and in the adiabatic approximation $\xi=vt$, i.e. the motion 
becomes linear because the
periodic nonlinearity effectively compensates exactly the periodic potential.

More sophisticated evolution scenarios can be observed for $k_2\neq k_1$ 
depending whether they are 
commensurable or not.

\section{Numerical Results}

We now proceed to describe our numerical results comparing with
the analytical prediction of the previous section. We use 
$\epsilon=0.1$ (for which we expect the perturbative description
to still be meaningful), and vary the relative parameters of the 
two lattices (linear and nonlinear).

\begin{figure}[tbp]
\begin{center}
\epsfxsize=6.0cm \epsffile{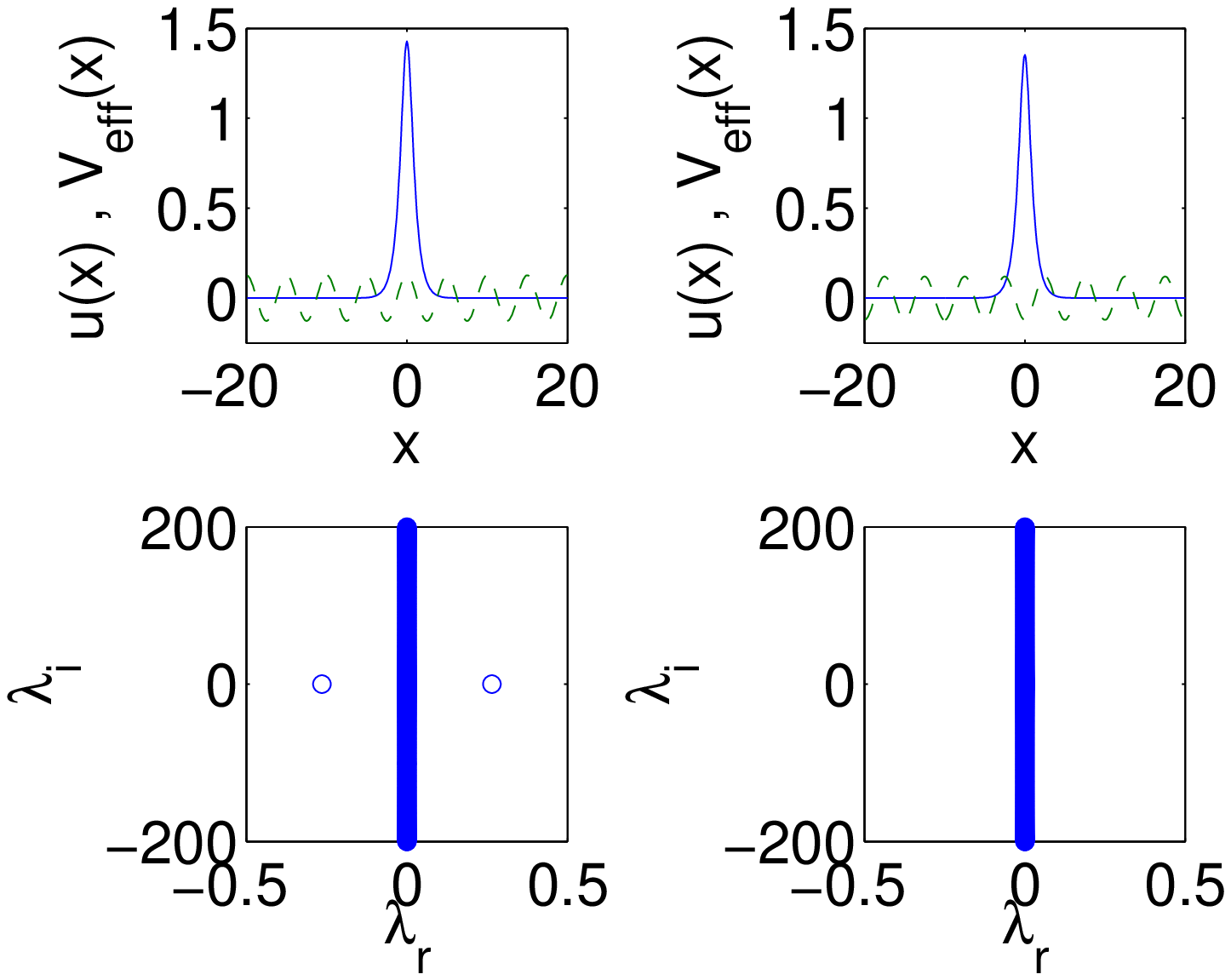} 
\epsfxsize=6.0cm\epsffile{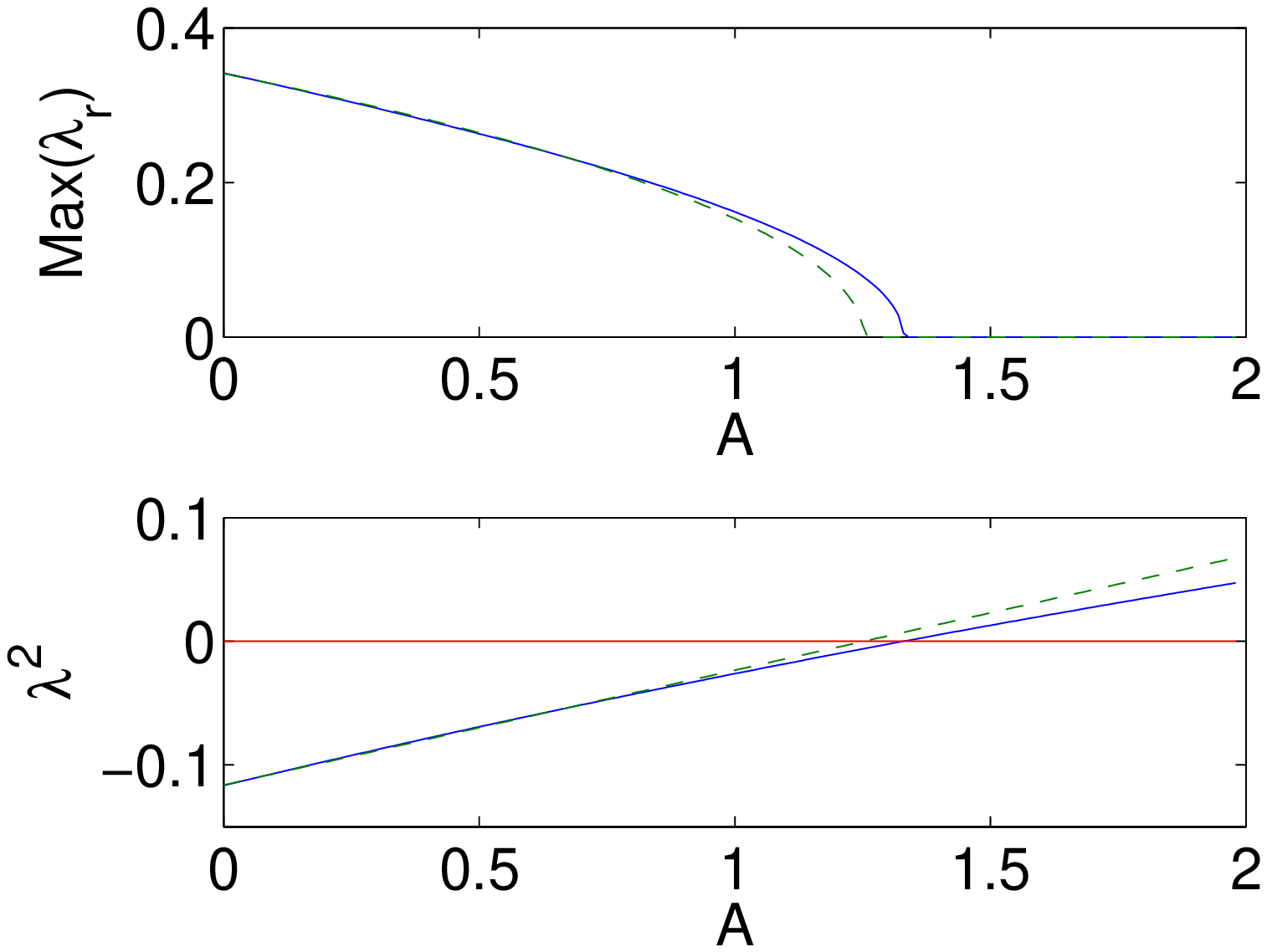} 
\epsfxsize=6.0cm\epsffile{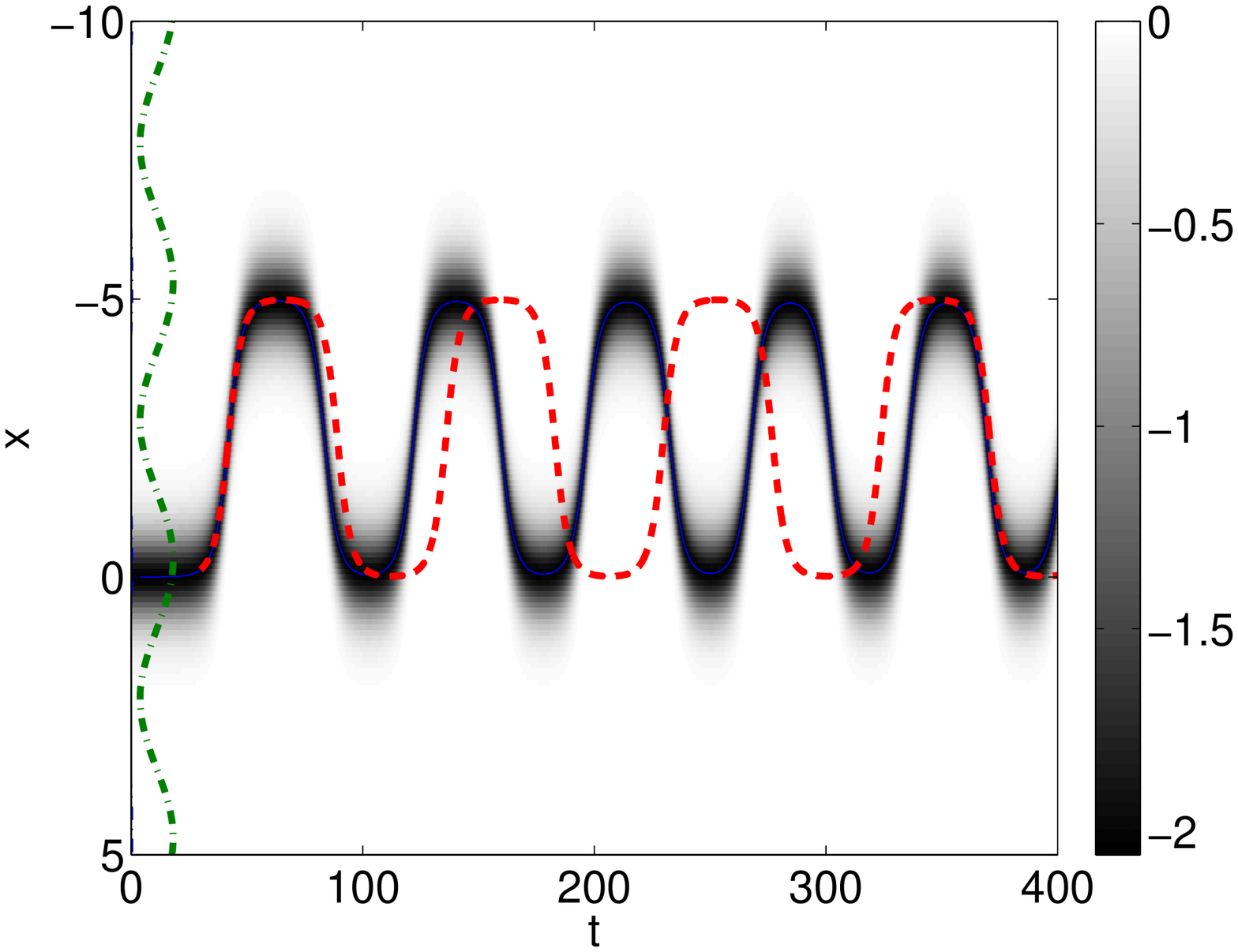} 
\caption{The top left set of panels shows the cases of $A=0.5$ (unstable,
left) and $A=2$ (stable, right). Both the solution profile (solid line)
and the effective potential (dashed line) is shown in the top subplots,
while the spectral plane of $(\lambda_r,\lambda_i)$ is shown in the bottom.
The presence of a real eigenvalue indicates instability in the latter. 
The right panels show the ``trajectory'' of the relevant eigenvalue
(real part in the top and squared eigenvalue in the bottom) as a function
of $A$. The solid line indicates the numerical result while the dashed line
the analytical prediction for the eigenvalue. The bottom panel shows
the evolution of the unstable configuration for $A=0.5$, in the effective
potential (shown out of scale in the graph by a dash-dotted line). In this 
spatio-temporal contour plot of the squared modulus (in fact, its opposite
is shown for clarity) of the solution, the result of the solitary dynamics
of Eq. (\ref{adiabatic}) is superposed as a thick dashed line.
It is clear that the solution, as a result of the instability, oscillates
between two maxima of the effective potential.}
\label{fig1}
\end{center}
\end{figure}

Our first set of numerical results consists of setting $B=1$ and $k_1=k_2=
2 \pi/5$, $\Delta \phi=0$ and varying $A$. In this way, we can
test the validity of our predictions for amplitude variations
(in this case of the nonlinear lattice). Our results are summarized
in Fig. \ref{fig1}. We have varied $A \in [0,2]$, finding that there
is a stability change within this interval. In particular, the left
panel of the figure shows the case of $A=0.5$ which is unstable,
and of the stable $A=2$. The right panel shows this transition in
terms of the real part (and also of the square) of the relevant
eigenvalue associated with the translational mode. It is found that 
this eigenvalue pair starts out as real, for small $A$, and 
becomes imaginary for $A>1.33$. We use Eq. (\ref{extra2}) 
to theoretically predict this transition as occurring at 
$A=1.253$. While we see that both qualitatively and fairly quantitatively
the dependence of the eigenvalue on the parameter is captured 
accurately by our theoretical result, it is meaningful to rationalize
the $\approx 6 \%$ error in the critical point estimation. It is,
in fact, observed that the soliton does not maintain its amplitude
in this continuation process (as a function of $A$), but rather
that its amplitude is reduced from $\approx 1.455$ for $A=0$ to
$1.353$ for $A=2$. This clearly shifts the critical point upwards,
whose analytical expression in this setting of $k_1=k_2$ and $\Delta
\phi=0$ can be easily seen to be $A_{cr}=12 B/(k^2+4 \mu)$. This
is in agreement (in fact, even quantitatively, if one uses the
above amplitude variation) with what is observed in our numerical
results.

In the bottom panel of the figure, we show the result of the unstable
dynamical evolution for the case of $A=0.5$. It can be seen that as
a result of the dynamical instability the solitary wave starts moving
to the left, eventually executing oscillations between the two 
maxima of the effective potential of Eq. (\ref{cintgeneral}).
In the same plot, we show the result of the adiabatic soliton perturbation
theory in this case (this is a rather ``stringent'' test of the theory
given the unstable dynamical evolution). We observe that the Eq. 
(\ref{adiabatic}) performs well in approximating the soliton trajectory
over the first oscillatory cycle. However, for longer times, we observe
it to gradually increasingly fail to capture the relevant oscillation.
This can be seen to be due to the fact that the solitonic trajectory
emits small wakes of radiation as it arrives at the turning points,
resulting in a weakly damped oscillation, a feature which is not captured by
our present considerations. However, we note in passing that methods 
similar to those developed by Soffer and Weinstein \cite{soffer}
can be used to rigorously account for such corrections.


\begin{figure}[tbp]
\begin{center}
\epsfxsize=6.0cm \epsffile{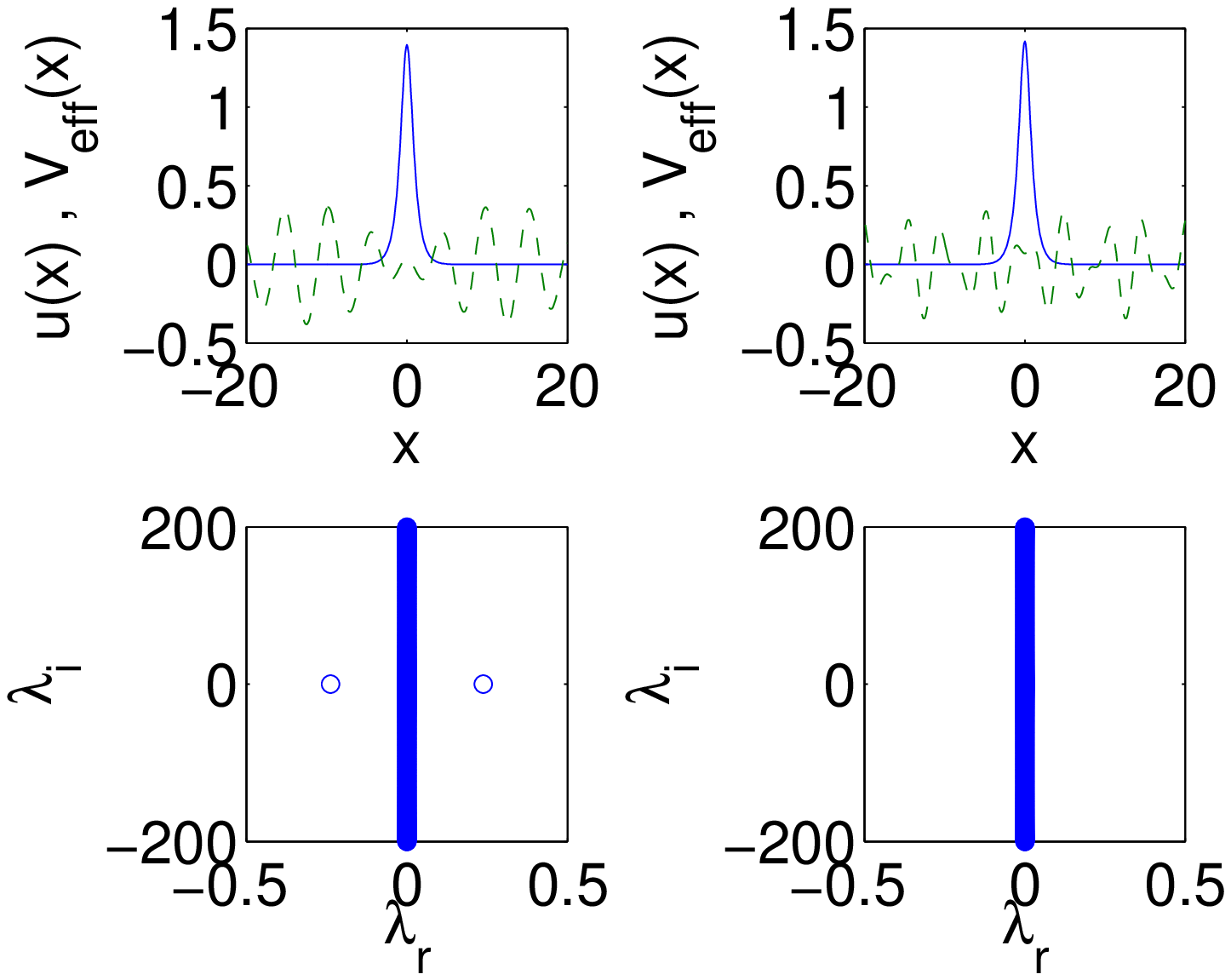} 
\epsfxsize=6.0cm\epsffile{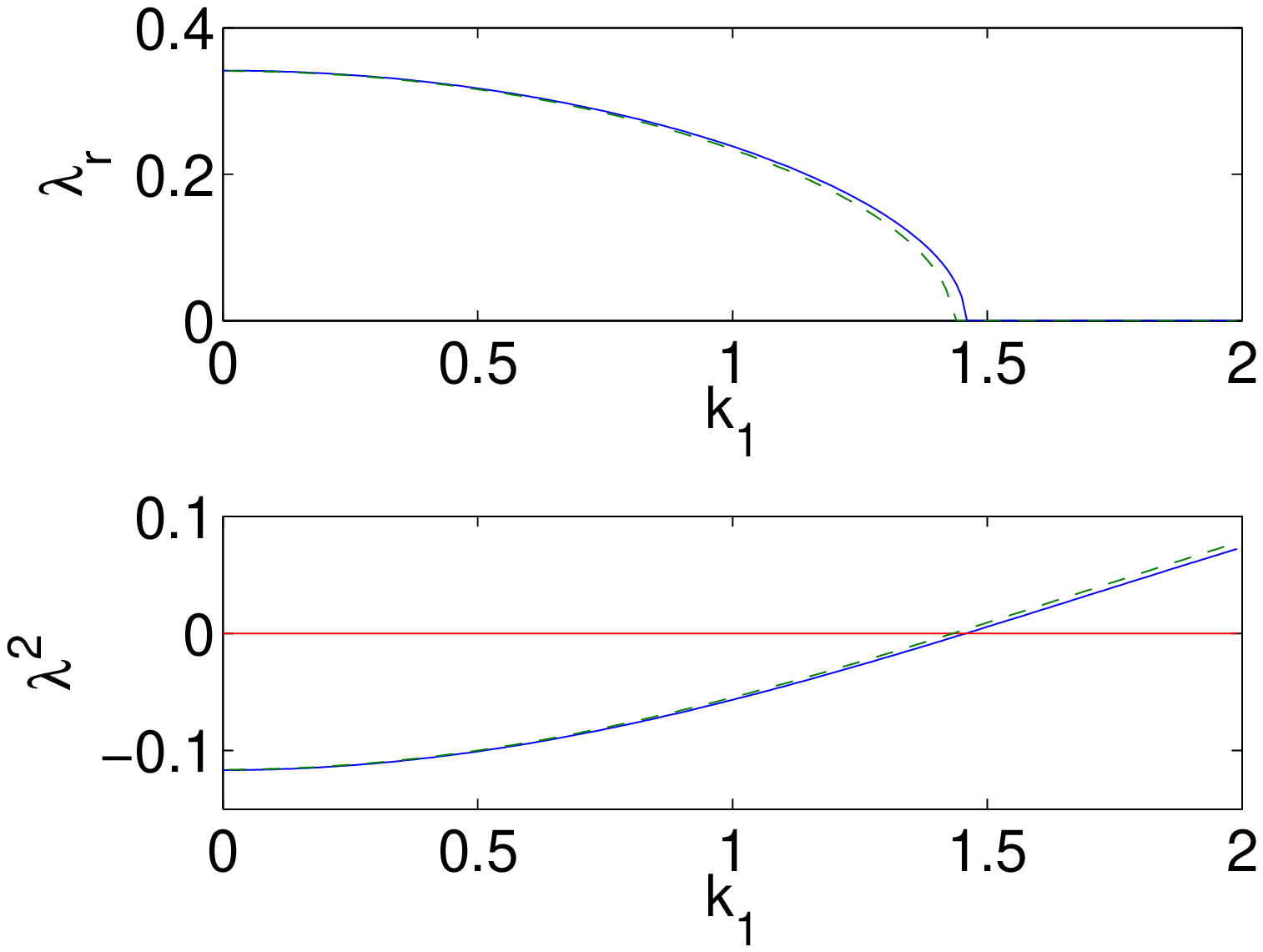}
\epsfxsize=6.0cm\epsffile{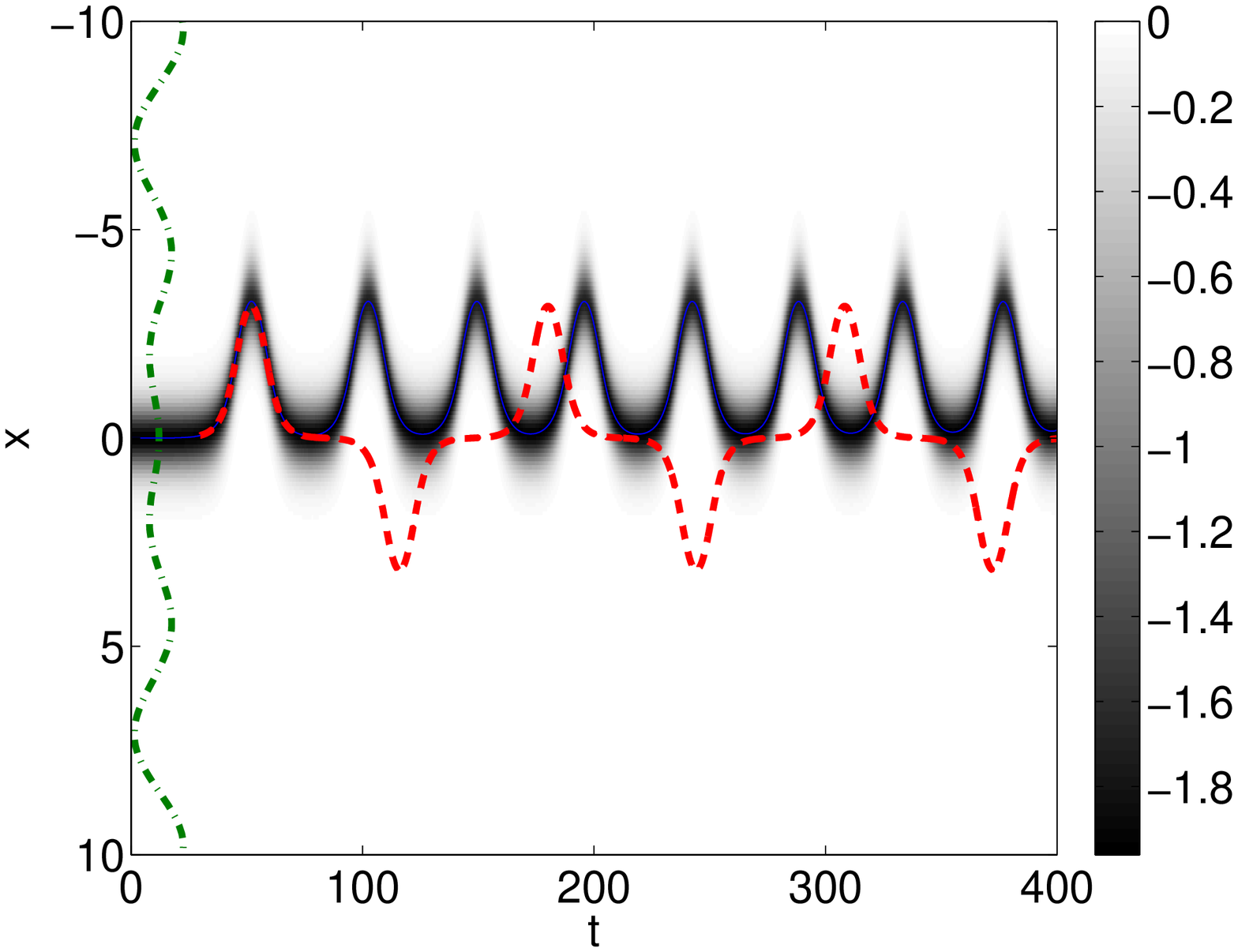}
\caption{Same as Fig. \ref{fig1}, but for the case where $k_1$
is varied (i.e., one of the wavenumbers, instead of the amplitude
$A$). The top left panels show the solution (solid line) and effective
potential (dashed line) for $k_1=1$ (left subplots) and $k_1=2$
(right subplots), as well as the corresponding spectral plane 
for the stability (bottom subplots). The top right panels show
the theoretically predicted (dashed line) versus numerically obtained
(solid line) eigenvalue of interest. The bottom panel shows the
spatio-temporal evolution of the unstable case with $k_1=1$, exhibiting
oscillations in the effective potential landscape (shown out of scale
by a dash-dotted line).}
\label{fig2}
\end{center}
\end{figure}

Our second parameter variation involved the role of the wavenumbers.
In particular, for the results reported in Fig. \ref{fig2}, we
have used $A=B=1$, and fixed $k_2=2 \pi/5$ and $\Delta \phi=0$, 
varying $k_1$. One can see that in this case, the effective potential
landscape changes significantly in term of its local structure
(in the previous example, it did not change, in that it was simply
two cosinusoidal terms with different signs, so it was simply a 
matter of which had the largest ``strength''). The other important
feature is that in this case, as well, there is a transition 
from instability to stability, as $k_1$ is increased. In fact,
the theoretical prediction for the critical point is $k_1=1.433$,
while the numerical one is $k_1=1.46$. Once again this can be seen for the 
two different settings of the left panel (the unstable case of
$k_1=1$ and the stable case of $k_1=2$), and is captured extremely
accurately by the prediction of Eq. (\ref{extra2}) about the
location of the relevant eigenvalue (associated with translation).
Furthermore, once again, the effective potential landscape that
can be computed from Eq. (\ref{cintgeneral}) can provide 
very useful information not only about the stability of local
extrema but also about the instability dynamics. The latter
is observed in the bottom panel of the figure.
Note, however, that while the effective potential predicts
accurately the turning points of the solitary wave dynamics,
the situation is more complicated with the dynamical equation
of motion of (\ref{adiabatic}). While, once again, the latter
predicts very accurately the first oscillation cycle, its non-accounting
of the radiative corrections of the motion leads to dynamics that
overcomes the shallow potential barrier at $x=0$; this is not
true, however, for the full PDE dynamics. This should serve as
a note of caution in regard to using the adiabatic approximation in such (marginal) cases.

We also explored the role of the phase difference 
between the linear and nonlinear lattice, by varying 
$\Delta \phi \in [0,2\pi]$, for $A=B=1$ and $k_1=k_2=2 \pi/5$. One of
the particularly interesting features of this example
is that while the instability of the original configuration
is not modified by this variation, the location of the 
solitary wave is. This is naturally expected on the basis
of Eq. (\ref{cintgeneral}). In particular, we observe that
the bright soliton's center location features an oscillation around
$\xi=0$, of period $ 2 \pi$ (as expected); for $\Delta \phi \in (0,\pi)$,
the wave is shifted to the right, while for $\Delta \phi \in (\pi,2 \pi)$,
it lies to the left of the origin. In this case, we examine both
the prediction for the relevant unstable eigenvalue, as well as
the prediction of our theoretical results for the location
of the center of the structure. The numerical results once again
align extremely well with the theoretical ones, confirming the
validity of our theoretical findings.

\begin{figure}[tbp]
\begin{center}
\epsfxsize=6.0cm \epsffile{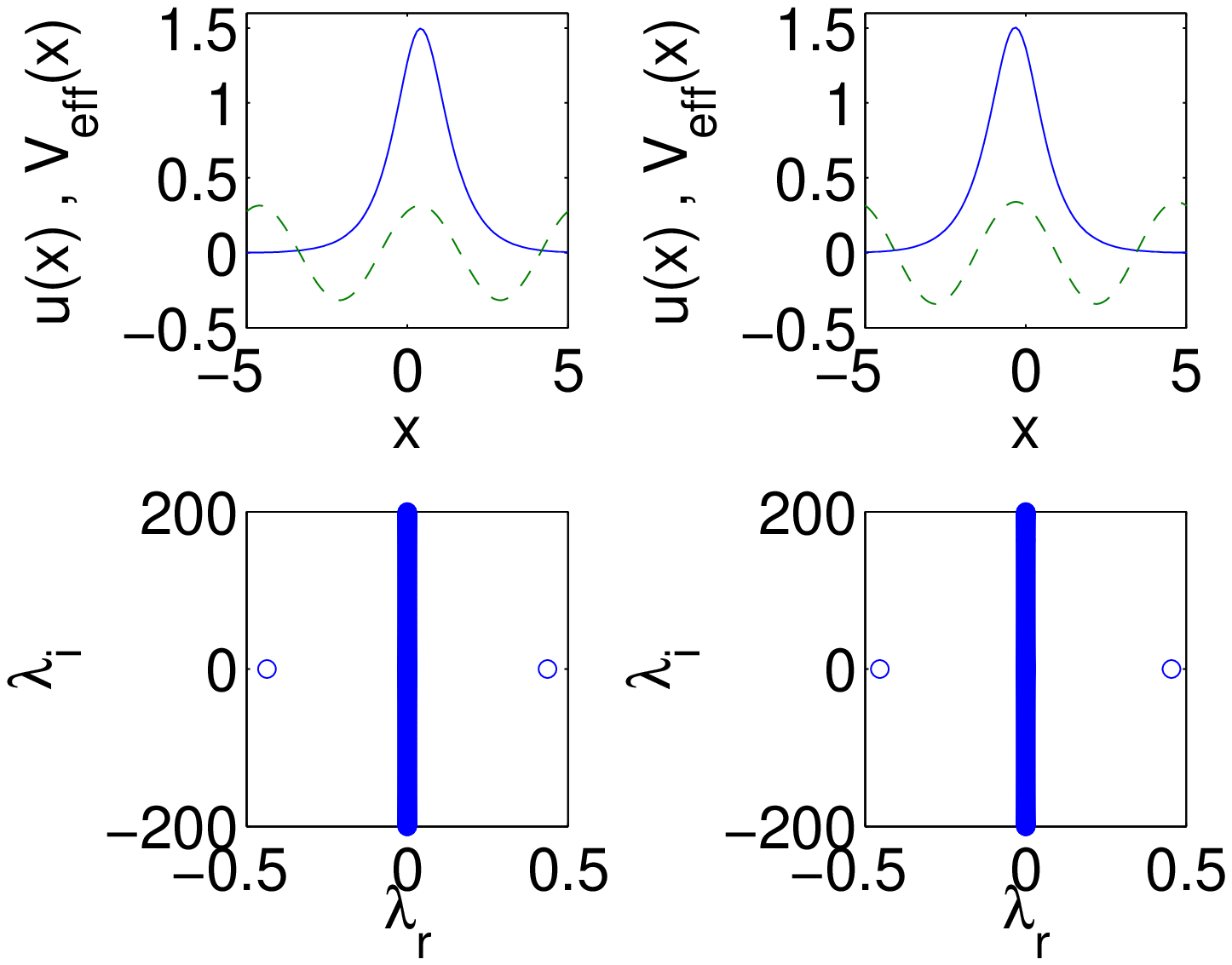} 
\epsfxsize=6.0cm\epsffile{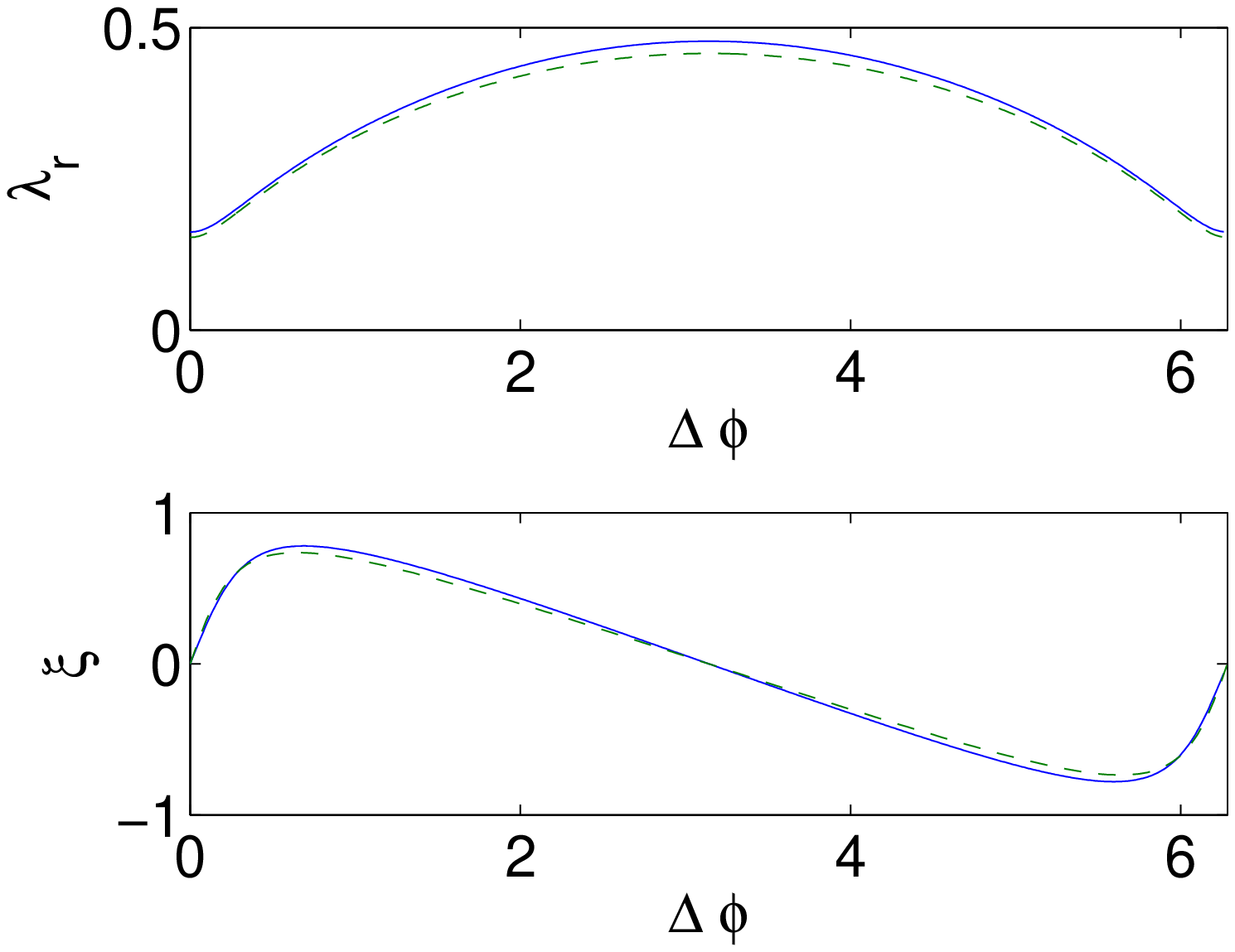} 
\caption{Same as the top panels of the previous figures. The left
panels show the configuration and its stability for $\Delta \phi=2$
(leftmost panel, the wave is shifted to the right) and $\Delta \phi=4$ 
(where the wave is shifted to the left). The right panel shows the
relevant unstable translational eigenvalue in the top subplot and
the center location of the wave in the bottom subplot. The theoretical
results are given by dashed lines and compare very favorably with the
solid lines of the full numerical results.}
\label{fig3}
\end{center}
\end{figure}

\begin{figure}[tbp]
\begin{center}
\epsfxsize=6.0cm \epsffile{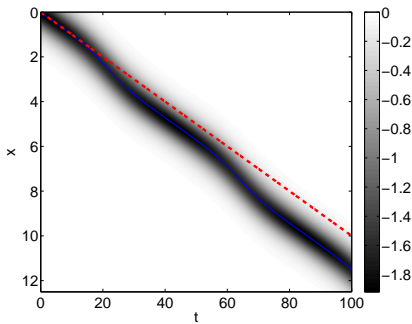} 
\caption{Space-time contour plot of the soliton evolution over a linear
and nonlinear lattice with $A=1.33$, $B=1$, $k_1=k_2=2 \pi/5$ and $\Delta 
\phi=0$. The thick dashed red line shows the curve $x=0.1 t$ for comparison.
Notice the very weak modulation in the near-free 
propagation of the solitary wave.}
\label{fig4}
\end{center}
\end{figure}

Finally, we also considered a case in the neighborhood of the complete
mutual cancellation of the two contributions of the effective potential.
In particular, for the case of $k_1=k_2=2 \pi/5$, and for $B=1$, $\Delta 
\phi=0$, we examined the dynamics for $A \approx 1.33$, whereby in
accordance with Fig. \ref{fig1}, the relevant (translational) 
eigenvalue is numerically found to be returning to $\lambda^2=0$, 
thereby restoring a regime of effective translational invariance.
In this setting, we initialized a stationary soliton boosted by
a factor of $\exp(i \kappa x)$, with $\kappa=0.1$, in Fig. \ref{fig4}
(other values of $\kappa$ were also used with similar results). 
We observe that the soliton appears to propagate with a speed near
the originally ``assigned'' one, being submitted only to very weak
modulations due to the very weak (in this case) effective potential.
These modulations are accompanied by oscillations of the solitary wave
amplitude and lead to a speed slightly larger than 0.1 (shown by thick
dashed line in Fig. \ref{fig4}). We note that in this way we can induce
the robust motion of the waves over the combined linear and nonlinear
lattice terrain.

\section{Conclusions} 

In this paper, we have examined
the evolution of bright solitary waves
in the presence (and competing effects) of
linear and nonlinear lattices. We have computed
the effective potential landscape that the
wave encounters and have explained how its
curvature is associated with the wave stability.
The relevant translational eigenvalue has been
explicitly computed and the transitions from
stability to instability due to the zero crossing
of this eigenvalue have been quantified as a
function of the system's parameters. 
The same threshold condition has been obtained 
independently from a direct eigenvalue calculation.
It has been shown that these theoretical
frameworks capture accurately the location
of the stationary waves, as well as the pertinent
eigenvalues, hence they constitute valuable tools for
inferring existence and stability information
about the coherent structures of such models. On the other hand,
the ensuing potential energy landscape can be used to derive a dynamical
equation for the motion of the soliton; however,
there is a number of notes of caution that should
be made in that regard, as the latter may not
capture entirely accurately the dynamical behavior,
especially in ``marginal'' cases, due to the role
of radiative corrections. It is also possible to
appropriately tune the system parameters so as to
nearly mutually cancel the effects of the linear
and nonlinear lattice and produce a wave that is
propagating at nearly-constant speed.

It would be especially interesting to apply
similar considerations to the case of dark
solitons in repulsive BECs 
and examine their impact on the spectrum
in the spirit of the recent work of \cite{pelinovsky}.
Similar considerations could subsequently be extended
to higher dimensional settings, where an important
relevant example would be the influence of the lattice 
potentials to the existence and stability of structures with vorticity
\cite{jpb,mplb2}.

PGK acknowledges support from
NSF-CAREER, NSF-DMS-0505663 and
NSF-DMS-0619492, as well as the
warm hospitality of MSRI during the
final stages of this work.
CKRTJ acknowledges the support of NSF-DMS-0410267 and the 
warm hospitality of MSRI.
VVK acknowledges support from Ministerio de Educaci\'on y Ciencia 
(MEC, Spain) under 
the grant SAB2005-0195 and support of FCT and FEDER under the grant 
POCI/FIS/56237/2004.

\end{document}